\documentclass[conf]{NFFDY}

\usepackage[utf8]{inputenc}
\usepackage{multirow}
\usepackage{comment}
\usepackage{graphicx}
\usepackage{subcaption}
\usepackage{amsmath}
\usepackage{amsmath}
\usepackage[version=4]{mhchem}
\usepackage{siunitx}
\usepackage{longtable,tabularx}
\title{Reduced order modelling of air puff test for corneal material characterisation}

\author{Osama M. Maklad\footnote{\href{mailto:O.Maklad@greenwich.ac.uk}{O.Maklad@greenwich.ac.uk}
} }
\affil{Centre for Advanced Manufacturing and Materials, University of Greenwich, London, UK}
\author{Muting Hao\footnote[3]{\href{mailto:Muting.Hao@eng.ox.ac.uk}{Muting.Hao@eng.ox.ac.uk}
}}
\affil{Oxford Thermofluid Institute, University of Oxford, Oxford, UK}

\begin{document}

\maketitle

\begin{center}
   \textbf{Abstract} 
\end{center}

\begin{abstract}

Models of the fluid-structure interaction (FSI) model for the air puff test were analysed. Using Abaqus, the air puff test is applied to eyes with varying biomechanical parameters, such as material properties, corneal thickness, and radius. A reduced order model of the air puff (a turbulent impinging jet) has been acquired to decrease simulation time from 48 hours for the FSI model to approximately 12 minutes for the finite element analysis (FEA) model alone. To further accelerate simulations and improve model accuracy, Physics-Informed Neural Networks (PINNs) will be integrated with the reduced-order model. This hybrid approach will help expand the model to a larger dataset, enhancing intraocular pressure (IOP) estimation accuracy and the corneal material properties algorithm through inverse FEA.
Additionally, a neural network (NN) framework with embedded Gaussian-modulated waveforms is proposed to model the pressure and deformation distributions on the corneal surface as functions of spatial and temporal parameters. By learning the relationship between corneal biomechanical inputs such as Corneal Central Thickness (CCT), Intraocular Pressure (IOP), and baseline properties ($\mu$), and the governing coefficients of pressure and deformation, the network accurately reconstructs the result that matches well with the high-fidelity CFD data. This approach can quickly capture the distribution of pressure and deformation. It can also provide insights into the distinct spatial and temporal dynamics of pressure and deformation, giving a more comprehensive understanding of fluid-structure interaction phenomena in the air puff test.

\end{abstract}


\textbf{Nomenclature}

{\renewcommand\arraystretch{1.0}
\noindent\begin{longtable*}{@{}l @{\quad=\quad} l@{}}
$IOP$  & Intraocular pressure \\
$CCT$ &    Central corneal thickness \\
$\mu$& Corneal stiffness parameter \\
$GBR$ & Gradient boosting regression \\

\end{longtable*}}

\section{Introduction}

In ophthalmology, the non-contact air puff tonometry test is widely used to measure the biomechanical properties of the human cornea and intraocular pressure (IOP). Accurate IOP measurement is critical in evaluating patients at risk of eye diseases such as glaucoma, where elevated IOP can lead to optic nerve damage. Glaucoma is one of the leading causes of asymptomatic permanent blindness in the developed world. A common cause of IOP increase is the improper drainage of aqueous humour due to trabecular meshwork blockage \cite{Weinreb2016}. The Ocular Response Analyzer (ORA) was the first tonometry device that utilized an air puff to assess ocular biomechanical properties, using a dynamic infrared signal analysis \cite{Luce2005}. Later, the Corvis-ST tonometer was developed, incorporating an ultra-high-speed Scheimpflug camera for better visualization of corneal deformation \cite{Eliasy2019}. The Corvis-ST applies a concentrated air puff to the centre of the cornea, deforming its geometry, which subsequently regains its original shape due to its elasticity. Using image processing, corneal deformation is recorded to estimate biomechanical properties and IOP through a programmed parametric equation.

Keratoconus, a degenerative eye disease, is characterized by the progressive thinning of the cornea, resulting in a cone-shaped protrusion \cite{Rabinowitz1998}. This condition significantly alters the cornea’s thickness, shape, and biomechanical properties \cite{Ambrosio2017}, often causing irregular astigmatism and blurry vision. In patients with keratoconus, tonometry measurements of IOP tend to be lower due to the strong correlation between IOP readings and the altered biomechanical properties, particularly corneal central thickness (CCT) \cite{McMonnies2016,Saad2012}. Understanding these biomechanical changes is crucial for clarifying the pathophysiology and aetiology of keratoconus, aiding in its treatment \cite{Roberts2017}.

Accurate in vivo measurements of both IOP and corneal biomechanical parameters are essential. However, the challenge lies in the mutual dependence between these parameters, making it difficult to isolate the effects of IOP from biomechanical properties like thickness and material stiffness on corneal response \cite{Eliasy2019}. Solving an inverse problem offers a potential solution to this challenge, allowing for a more accurate assessment of corneal material behaviour based on improved IOP values \cite{maklad2020fsi,He2018}.

The study of fluid-structure interaction (FSI) between the air puff and the cornea is necessary to address the correlation between IOP and corneal parameters and improve IOP estimation accuracy \cite{Sun2020}. To achieve this, the air puff pressure distribution profiles on the cornea must be considered when measuring corneal deformation \cite{Eliasy2019}. The pressure exerted on the cornea dynamically changes in response to corneal deformation, which can alter clinical interpretations \cite{Yousefi2018}. This effect termed the Corneal Load Alteration with Surface Shape (CLASS), highlights the importance of obtaining patient-specific air puff pressure distributions \cite{Yousefi2018,Eliasy2019,maklad2020fsi}.

While the co-simulation of FSI has provided accurate values for pressure distribution and corneal deformation \cite{maklad2020fsi,Chen2016,SinhaRoy2017}, it remains time-consuming. The evolution of FSI approaches has led to precise estimations of air jet pressure changes with corneal characteristics, resulting in biomechanically corrected equations for accurate IOP and corneal material stress-strain indices \cite{Eliasy2019,maklad2020fsi,Yousefi2018}. However, predicting air puff pressure in the computational fluid dynamics (CFD) model of FSI is computationally expensive, taking considerable time for patient-specific geometries.

The primary goal of this study is to investigate how individual corneal characteristics affect air puff pressure measurements and develop a machine learning (ML) algorithm to reduce computational costs. While numerical models of fluid flow are central to research on physical and mechanical phenomena, especially those involving FSI \cite{maklad2020fsi,Yousefi2018,Chen2016}, recent studies \cite{Liu2019,Gultepe2020,Zhu2021} suggest that ML can replace time-consuming numerical solvers, providing reduced-order models, improved optimization, and lower computational costs. Therefore, we have developed a supervised regression ML algorithm using the Gradient Boosting Regressor (GBR) to estimate time-dependent air puff corneal pressure distribution profiles, replacing the CFD model in the FSI co-simulation of \cite{maklad2020fsi} to reduce computational costs. This study focuses on the effect of changing corneal parameters (IOP, CCT, and material stiffness) on air puff pressure profiles, proposing a GBR-based algorithm for estimating patient-specific air puff pressure loads, and applying the predicted loads to the finite element model of the eye in ABAQUS software to simulate corneal deformation without the need for a time-consuming CFD model.

To further improve the efficiency and accuracy of corneal response predictions, this study introduces a hybrid Neural Network (NN)with parametric Embedding framework to model dynamic pressure and deformation distributions on the corneal surface. Observed patterns of pressure and deformation resemble Gaussian-modulated waveforms \cite{Hassan2010}, motivating the use of a parametric model characterized by five coefficients: amplitude ($A$), spatial scaling ($\beta$), temporal characteristics ($\mu$ and $\sigma$), and spatial attenuation ($\alpha$).
In this study, the NN is trained to map biomechanical input parameters—Corneal Central Thickness (CCT), Intraocular Pressure (IOP), and baseline material properties ($\mu_0$)—to these coefficients, enabling efficient reconstruction of pressure and deformation distributions. This approach accelerates computational workflows, replacing the need for expensive CFD simulations. It can also provide new insights into the spatial and temporal dynamics of corneal responses.

\section{Method}

\begin{figure}[hbt!]
\centering
\includegraphics[width=\textwidth]{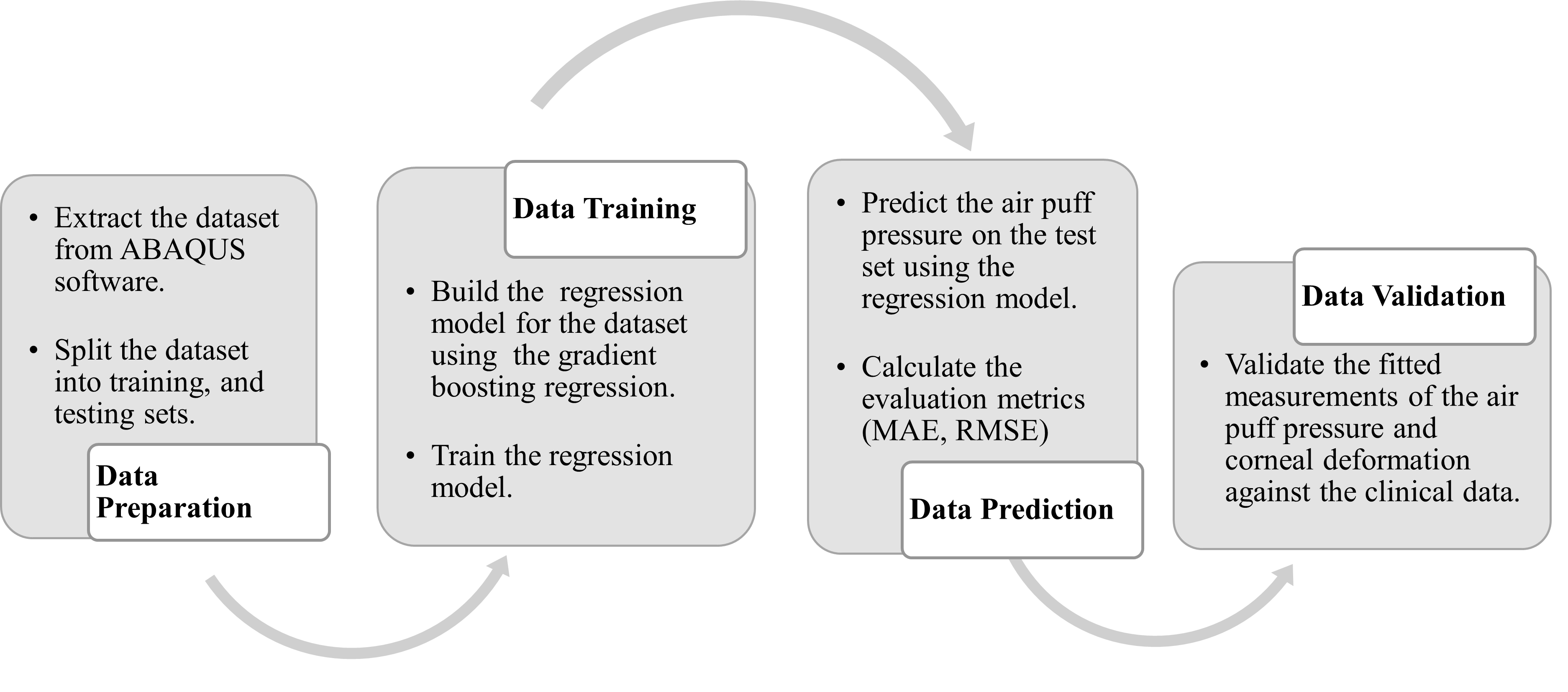}
\caption{Flowchart of the data processing in the current study.}
\label{fig1}
\end{figure}

This section outlines the dataset used in our approach and describes the algorithm applied for learning and prediction. We then explain the evaluation metrics used to assess the model's performance and conclude the section by presenting a validation study comparing the clinical data from patient cases with the outcomes of our algorithm. The data processing steps are illustrated in Figure \ref{fig1}, which will be elaborated upon in the subsequent sub-sections.

\subsection{Data Collection and Processing } 

\begin{figure}[hbt!]
\centering
\includegraphics[width=\textwidth]{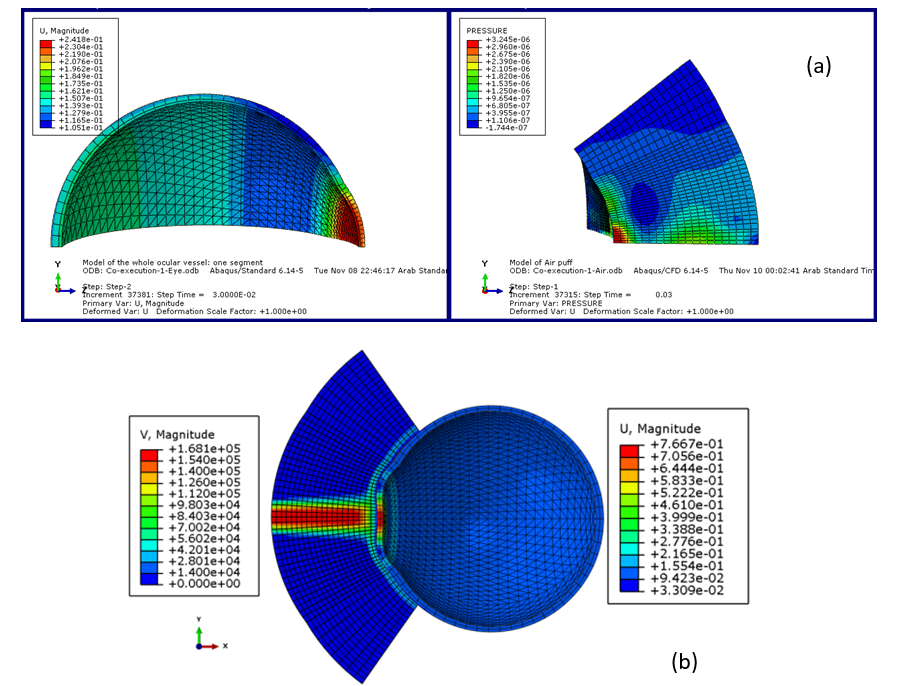}
\caption{The FSI coupled model of the air puff test in ABAQUS software \cite{maklad2020fsi}.}
\label{fig2}
\end{figure}

This section details the dataset used as input to our algorithm, which was derived from the model of \cite{maklad2020fsi}, involving a novel multi-physics, fully coupled Fluid-Structure Interaction (FSI) model of the air-puff test of the Corvis ST on eye globes subjected to the internal load of IOP as shown in Figure \ref{fig2}. 
Details of the numerical model, including, its validation and the used FSI two-way coupling approach with all the co-simulation control parameters and equations were published in our earlier study \cite{maklad2020fsi}. Here, we are giving the most important information, the air-puff was simulated using the turbulent Abaqus/CFD solver (version 6.14-2, Dassault Systèms Simulia Inc., USA) coupled with the finite element model of the eye using an arbitrary Lagrangian-Eulerian (ALE) deforming mesh. Models of the air domain consisted of six-nodded 3D fluid continuum elements (FC3D6) and used Spalart–Allmaras turbulent eddy viscosity model to simulate the turbulence in the air jet. To avoid excessive distortion of the air domain mesh during the coupling process with the eye model, an adaptive Arbitrary Lagrangian–Eulerian (ALE) deforming mesh was used to improve the stability of the simulation analysis.
Due to the rotational symmetry of the parametric study results by \cite{maklad2020fsi} in both domains, a quarter of the two domains was simulated to reduce computational time, as illustrated in Figure \ref{fig2}(a). Figure \ref{fig2}(b) shows the deformation values of the entire ocular vessel. The air puff velocity values are indicated on the left (in mm/s), while the deformation values of the eye model are shown on the right (in mm).

This study considered three primary parameters: intraocular pressure (IOP), central corneal thickness (CCT), and the material stiffness coefficient ($\mu$). The selection of five values for the material stiffness coefficient was based on the age-related relationship established by \cite{elsheikh2010age}, where $\mu$=0.0328 corresponds to an age of 30 years and $\mu$=0.1082 corresponds to an age of 100 years. The effect of each parameter on the corneal pressure distribution was analyzed while keeping the other parameters constant.

The dataset used to develop the regression model comprised 17 numerical simulations generated using the FSI model in ABAQUS 6.14 \cite{maklad2020fsi}, along with four clinical cases used for validation. The dataset contained both pressure distributions and corneal deformations, providing a comprehensive basis for training and evaluating the model.

The numerical simulations were structured as follows:

\begin{itemize}
    \item First, the corneal pressure distribution was estimated for IOP values ranging from 10 to 25 mmHg, with a fixed CCT of 445 $\mu$m and a material stiffness coefficient of 0.0541.

    \item Next, the effect of varying CCT values (445, 495, 545, 595, and 645 $\mu$m) on the corneal pressure distribution was examined while maintaining an IOP of 15 mmHg and a material stiffness coefficient of 0.0541.

    \item Finally, the influence of different material stiffness coefficients (0.0328, 0.0541, 0.0683, 0.0811, and 0.1082) was analyzed while holding the IOP at 15 mmHg and the CCT at 545 $\mu$m.
\end{itemize}
For each experiment, the dataset was divided into 70\% for training, 20\% for testing, and 10\% for validation, ensuring robust model evaluation. The four clinical cases were not included in the training phase but were used separately for overall validation. The model was trained to estimate the pressure distribution on the corneal surface, with accuracy evaluated using the Mean Absolute Error (MAE) metric, which quantifies the average absolute difference between the observed and predicted pressure values.

A summary of the corneal parameters used in the numerical dataset is presented in Table \ref{table1}.

\begin{table}
\centering

\begin{tabular}{| l  |l  |l  |l  |l  |l  |l  |l  |l |}
\hline
IOP (mmHg) & 10 & 13 & 15 & 17 & 20 & 22 & 24 & 25 \\
\hline
CCT ($\mu$m) & 445 & 495 & 545 & 595 & 645 & \multicolumn{3}{|l|}{} \\ \hline 
$\mu$ & 0.0328 & 0.0541 & 0.0683 & 0.0811 & 0.1082 & \multicolumn{3}{|l|}{} \\ \hline

\end{tabular}

\caption{The values of the corneal parameters used to test our regression model. Please note that the number of parameters doesn't have to be equal. We used 8 increments for IOP as the results are more sensitive to the change in IOP}
\label{table1}
\end{table}

\subsection{The Gradient Boosting Regressor (GBR) Algorithm}

Gradient Boosting Machine (GBM) is a widely used machine learning algorithm for both regression and classification tasks. When applied to regression problems, this algorithm is referred to as the Gradient Boosting Regressor (GBR). The GBR approach relies primarily on three components: the loss function, the base learner, and an additive model structure \cite{friedman2001greedy}. In GBR, a series of tree-based models are constructed sequentially, with each subsequent model learning from the residuals (errors) of the previous one. By "boosting" these weak learners—typically decision trees—into an ensemble, GBR creates a more robust and accurate predictive model \cite{friedman2000additive}.

To effectively implement the GBR algorithm, several hyperparameters must be defined, as they significantly influence the model's performance and accuracy. In this study, we set the parameters as follows: squared-loss function, learning rate of 0.3, 2700 boosting stages, maximum tree depth of 6, and minimum samples required to split a node set to 5. The learning rate, a crucial parameter, determines how much each tree's prediction affects the final model. A lower learning rate slows the learning process but can enhance the model's stability and reliability.

The number of estimators (boosting stages) also plays an important role; generally, a higher number improves performance. The maximum depth parameter limits the depth of each decision tree, helping to prevent overfitting, while the minimum number of samples required to split a node influences the tree's branching structure \cite{scikit_gbr}.

All computations were performed on an Intel Core i7 8550U processor with 8 GB RAM. Below, we detail the mathematical formulation of the GBR model.

Given an input \( x \) with prediction \( \hat{y} \), the additive model of GBR can be formulated as follows \cite{scikit_gbr}:

\[
\hat{y} = \sum_{m=1}^{M} \gamma_m f_m(x)
\]

where \( M \) represents the `n\_estimators` parameter, \( \gamma_m \) is the learning rate, and \( f_m(x) \) is the base estimator (a weak learner).

The GBR algorithm is built iteratively, updating the model at each step. The updated model is given by:

\[
\hat{y}^{(m+1)} = \hat{y}^{(m)} + \gamma_m f_m(x)
\]

where each new tree \( f_m(x) \) is fitted to minimize the sum of losses \( L \) over the previous ensemble \cite{scikit_gbr}:

\[
L = \sum_{i=1}^{n} \ell(y_i, \hat{y}_i^{(m)})
\]

Thus, for a given loss function \( \ell \), the model equation becomes:

\[
\hat{y} = \hat{y}^{(0)} + \sum_{m=1}^{M} \gamma_m f_m(x)
\]

By default, for a least-squares loss function, the initial prediction \( \hat{y}^{(0)} \) is set to the mean of the target values. Using a first-order Taylor approximation, the incremental adjustment \( f_m(x) \) can be approximated as \cite{scikit_gbr}:

\[
f_m(x) \approx -\frac{\partial L}{\partial \hat{y}^{(m)}}
\]

where \( -\frac{\partial L}{\partial \hat{y}^{(m)}} \) represents the negative gradient of the loss function. After removing constant terms, this results in:

\[
f_m(x) = -\frac{\partial \ell(y, \hat{y})}{\partial \hat{y}}
\]

The algorithm continues iterating and updating gradients until it converges. This process can be seen as a form of gradient descent within a functional space \cite{scikit_gbr}.

To evaluate the performance of the GBR model, we measured model error using Mean Absolute Error (MAE) and Root Mean Square Error (RMSE). The MAE, representing the average absolute difference between observed and predicted pressure values, is given by:

\[
\text{MAE} = \frac{1}{n} \sum_{i=1}^{n} |y_i - \hat{y}_i|
\]

Similarly, RMSE, representing the standard deviation of prediction errors, is calculated as:

\[
\text{RMSE} = \sqrt{\frac{1}{n} \sum_{i=1}^{n} (y_i - \hat{y}_i)^2}
\]

Additionally, we considered the computational time required for the model to train and generate predictions as an important factor due to our focus on reducing computational costs.

\subsection{Validation of the GBR algorithm }

To validate our algorithm using clinical data, we utilized a diverse clinical dataset encompassing a range of corneal parameters from four healthy patients. This data was provided by the Vincieye Clinic in Milan, Italy, and the Rio de Janeiro Corneal Tomography and Biomechanics Study Group, Brazil. These clinical cases served as the input for the GBR algorithm, which was used to predict air puff pressure loading and to compare with observed corneal deformations. Compliance with the ethical guidelines outlined in the Declaration of Helsinki (1964) and its 2013 revision was maintained, and all patients provided informed consent prior to the anonymized use of their data in this research. Details of the clinical cases included are summarized in Table \ref{table2}.

\begin{table}
\centering

\begin{tabular}{| l | l | l | l | l |}
\hline
Corneal parameter/Case ID & Case 1

(Age=73) & Case 2

(Age=54) & Case 3

(Age=63) & Case 4

(Age=40) \\
\hline
IOP (mmHg) & 17.5 & 18 & 15 & 24 \\
\hline
CCT ($\mu$m) & 560 & 579 & 548 & 582 \\
\hline
$\mu$ & 0.061 & 0.054 & 0.057 & 0.051 \\
\hline

\end{tabular}
\caption{The corneal parameters of the four clinical cases used in the validation.}
\label{table2}
\end{table}

\begin{figure}[hbt!]
\centering
\includegraphics[width=\textwidth]{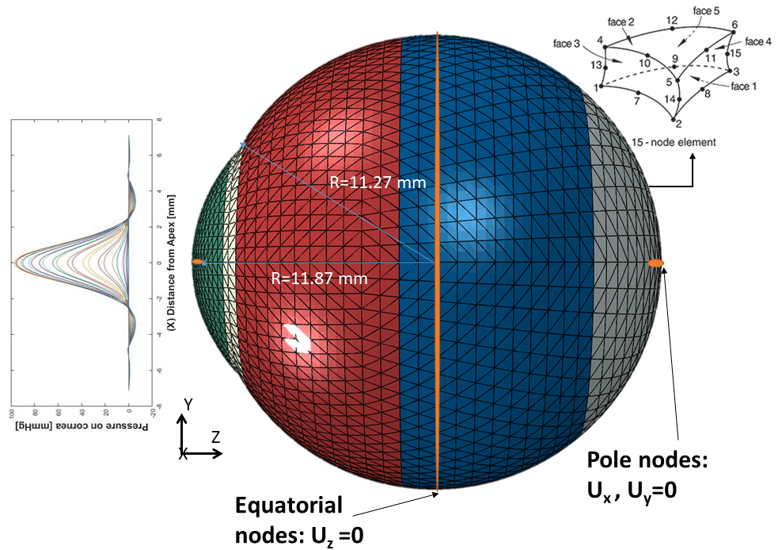}
\caption{The Finite Element model of the eye with the boundary conditions.}
\label{fig3}
\end{figure}

Once the air puff pressure loadings were predicted with the GBR algorithm, they were applied to the finite element (FE) model of the eye in ABAQUS 6-14. This three-dimensional eye model comprised 10,000 C3D15H continuum elements, each with fifteen nodes and nine integration points. These elements were arranged in two layers, spanning 15 rings along the cornea and 35 rings along the sclera. To stabilize the FE model, rigid body motion was restricted in the \(Z\)-direction at the equatorial nodes, while the posterior and anterior pole nodes were constrained in the \(X\) and \(Y\) directions, allowing free movement along the \(Z\)-axis, as illustrated in Figure \ref{fig3}. This configuration mirrors the setup used and validated in prior studies by the first author \cite{maklad2019, maklad2020a, eliasy2019ssi}. 

\subsection{A hybrid Neural Network (NN) with parametric Embedding framework of Gaussian-modulated waveforms}


\subsubsection{Overview of the Neural Network Framework}
This section proposes a neural network (NN) framework to model the pressure and deformation distributions on the corneal surface. Inspired by observed Gaussian-modulated wave patterns in experimental data, a parametric function was used to characterize the dynamics of pressure and deformation. This function, defined by five coefficients—amplitude ($A$), spatial scaling ($\beta$), temporal characteristics ($\mu$ and $\sigma$), and spatial attenuation ($\alpha$)—forms the basis for reconstructing pressure and deformation distributions. The NN is trained to map biomechanical inputs, including Corneal Central Thickness (CCT), Intraocular Pressure (IOP), and baseline material properties ($\mu_0$), to these coefficients. 

\subsubsection{Parametric Function for Pressure and Deformation}
The dynamic behavior of pressure and deformation is modeled using a Gaussian-modulated wave function, expressed as:
\begin{equation}
p_{\text{pred}}, dz_{\text{pred}} = A \cdot (1 - \alpha \cdot \beta \cdot x^2) \cdot \exp(-\beta \cdot x^2) \cdot \exp\left(-\frac{ (t - \mu)^2}{2 \sigma} \right)\label{eq:wavejet}
\end{equation}
where $p_{\text{pred}}$,  $dz_{\text{pred}}$ represents the predicted pressure or deformation using NN at cornea spatial position coordinate $x$ and time $t$. Five coefficients are additionally included:  $A$ is the amplitude of the pressure or deformation wave, $\alpha$ is the coefficient controlling spatial attenuation, $\beta$ is the spatial scaling parameter, $\mu$ is the temporal mean indicating the center of the wave and $\sigma$ is the temporal standard deviation, determining the wave spread. These coefficients capture both spatial and temporal dynamics and support the reconstruction of the distributions. $p_{\text{pred}}$ and $dz_{\text{pred}}$ follow the same function as Eq. \ref{eq:wavejet} but with different values of coefficients.

\subsubsection{Neural Network Architecture}
The NN framework is designed to map input parameters ($\text{CCT}$, $\text{IOP}$, $\mu_0$) to the output coefficients ($A$, $\alpha$, $\beta$, $\mu$, $\sigma$). Table~\ref{tab:nn_architecture} summarizes the network architecture.
The input layer have three nodes representing $\text{CCT}$, $\text{IOP}$, and $\mu_0$. Hidden Layers are two fully connected layers with 64 neurons each, employing ReLU (Rectified Linear Unit) activation to introduce non-linearity. The output layer includes five neurons, each representing one of the parametric coefficients ($A$, $\alpha$, $\beta$, $\mu$, $\sigma$). A linear activation function is used for the output layer to allow unrestricted values.

\begin{table}[h!]
    \centering
    \caption{Neural Network Architecture}
    \label{tab:nn_architecture}
    \begin{tabular}{|l|c|c|}
    \hline
    Layer             & Number of Neurons & Activation Function \\ \hline
    Input Layer        & 3                 & -                   \\ \hline
    Hidden Layer 1     & 64                & ReLU                \\ \hline
    Hidden Layer 2     & 64                & ReLU                \\ \hline
    Output Layer       & 5                 & Linear              \\ \hline
    \end{tabular}
\end{table}

\subsubsection{Loss Function}
The loss function is designed to minimize the Mean Squared Error (MSE) between the predicted and actual pressure or deformation values. Taking pressure for example:
\begin{equation}
\text{Loss} = \frac{1}{N} \sum_{i=1}^N \left(p_{\text{pred},i} - p_{\text{true},i}\right)^2,
\end{equation}
where $N$ is the total number of samples, $p_{\text{pred},i}$ is the predicted pressure, and $p_{\text{true},i}$ is the ground truth.

\subsubsection{Gradient Derivations for Backpropagation}
To update the NN weights using backpropagation, the gradient of the loss function with respect to each parametric coefficient is explicitly derived as follows:
    \begin{equation}
    \frac{\partial \text{Loss}}{\partial A} = \frac{2}{N} \sum_{i=1}^N (p_{\text{pred},i} - p_{\text{true},i}) \cdot (1 - \alpha \cdot \beta \cdot x^2) \cdot \exp(-\beta \cdot x^2) \cdot \exp\left(-\frac{(t - \mu)^2}{2 \sigma^2}\right).
    \end{equation}

    \begin{equation}
    \frac{\partial \text{Loss}}{\partial \alpha} = -\frac{2}{N} \sum_{i=1}^N (p_{\text{pred},i} - p_{\text{true},i}) \cdot A \cdot \beta \cdot x^2 \cdot \exp(-\beta \cdot x^2) \cdot \exp\left(-\frac{(t - \mu)^2}{2 \sigma^2}\right).
    \end{equation}

    \begin{equation}
    \frac{\partial \text{Loss}}{\partial \beta} = \frac{2}{N} \sum_{i=1}^N (p_{\text{pred},i} - p_{\text{true},i}) \cdot A \cdot \left[(\alpha \cdot x^2 - x^2) \cdot \exp(-\beta \cdot x^2)\right] \cdot \exp\left(-\frac{(t - \mu)^2}{2 \sigma^2}\right).
    \end{equation}

    \begin{equation}
    \frac{\partial \text{Loss}}{\partial \mu} = \frac{2}{N} \sum_{i=1}^N (p_{\text{pred},i} - p_{\text{true},i}) \cdot A \cdot (1 - \alpha \cdot \beta \cdot x^2) \cdot \exp(-\beta \cdot x^2) \cdot \frac{t - \mu}{\sigma^2} \cdot \exp\left(-\frac{(t - \mu)^2}{2 \sigma^2}\right).
    \end{equation}

    \begin{equation}
    \frac{\partial \text{Loss}}{\partial \sigma} = \frac{2}{N} \sum_{i=1}^N (p_{\text{pred},i} - p_{\text{true},i}) \cdot A \cdot (1 - \alpha \cdot \beta \cdot x^2) \cdot \exp(-\beta \cdot x^2) \cdot \frac{(t - \mu)^2}{\sigma^3} \cdot \exp\left(-\frac{(t - \mu)^2}{2 \sigma^2}\right).
    \end{equation}

\subsubsection{Training Configuration}
Each data point in the dataset corresponds to a unique combination of spatial position ($x$) and time ($t$), treated as a separate instance with its associated input parameters: Corneal Central Thickness (CCT), Intraocular Pressure (IOP), and baseline material property ($\mu_0$). These three parameters serve as the inputs to the neural network. 


The NN was optimized using the Adam optimizer, with an initial learning rate of $0.0005$ and decay rate $1e-3$.
The model was trained for 10000 epochs.

\section{Results}

\subsection{Effect of the corneal parameters on the air puff pressure }

To examine the impact of corneal parameters on the air puff pressure distribution, we assessed the Gradient Boosting Regressor (GBR) model's performance in predicting changes in corneal pressure distribution under varying conditions. Initially, the effect of intraocular pressure (IOP) variations on estimated pressure distribution was evaluated. Pressure distribution on the corneal surface was obtained for eight IOP values (10, 13, 15, 17, 20, 22, 24, and 25 mmHg), while keeping the central corneal thickness (CCT) constant at 445 $\mu$m and a material stiffness coefficient set at 0.0541. The fitted GBR model demonstrated strong alignment with the numerical values from the fluid-structure interaction (FSI) model, yielding a mean absolute error (MAE) of 0.0212, root mean square error (RMSE) of 0.0682, and an execution time of 12 seconds.

Next, the effect of CCT variations (445, 495, 545, 595, and 645 $\mu$m) on corneal pressure distribution was assessed at an IOP of 15 mmHg and a material stiffness coefficient of 0.0541. The GBR algorithm again aligned closely with the ABAQUS numerical model, with MAE = 0.0171, RMSE = 0.0578, and an execution time of 10 seconds. Finally, the influence of material stiffness coefficient variations, representing age-related stiffness changes (0.0328, 0.0541, 0.0683, 0.0811, and 0.1082), was examined at an IOP of 15 mmHg and CCT of 545 $\mu$m. The GBR model produced highly accurate pressure distributions, with an MAE of 0.0113, RMSE of 0.0491, and an execution time of 12 seconds.

\begin{figure}[hbt!]
\centering
\includegraphics[width=0.75\textwidth]{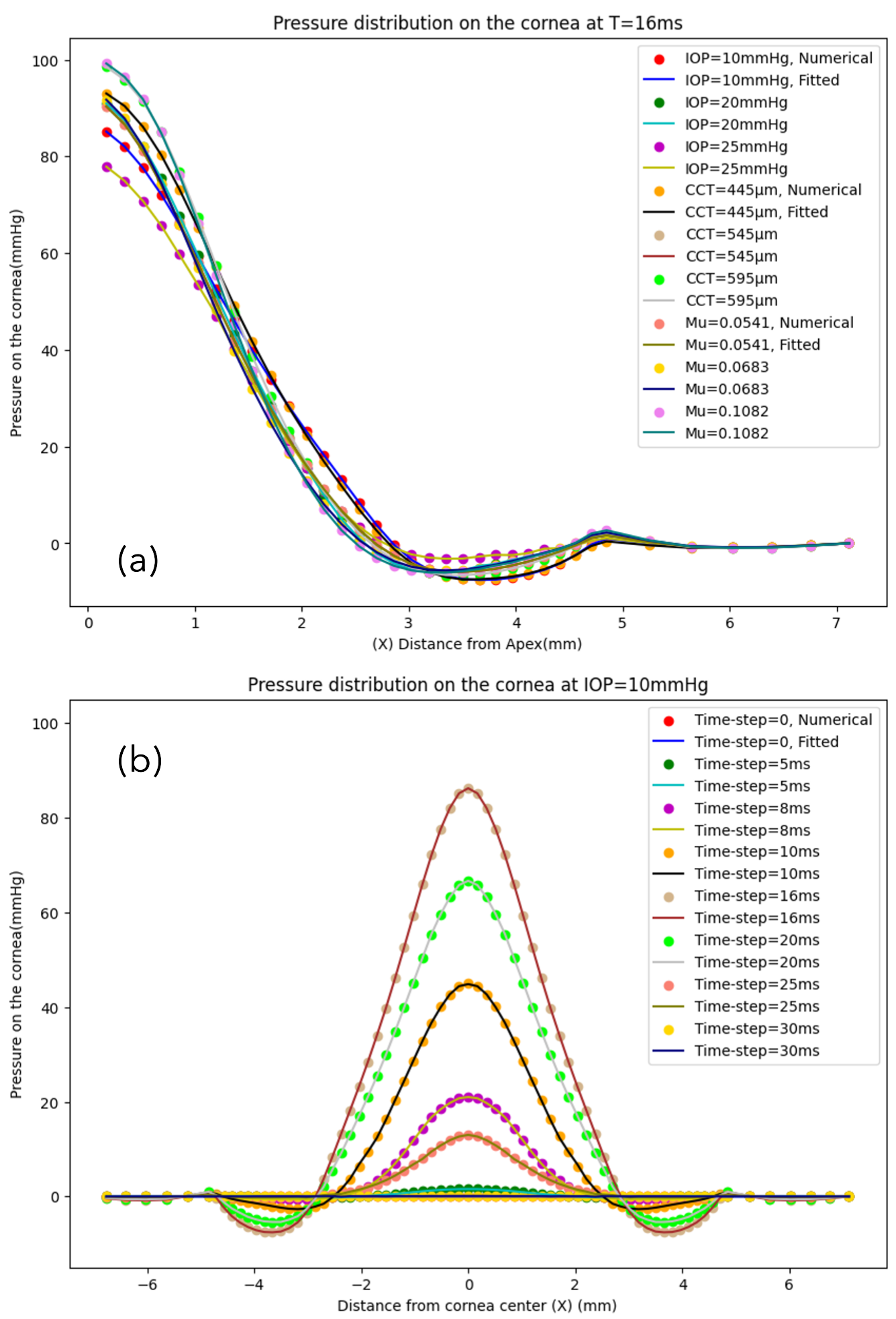}
\caption{(a)The comparison between the numerical ABAQUS model and the fitted GBR algorithm of the pressure distribution on the cornea at T=16ms with a different corneal parameter of IOP, CCT, and $\mu$, (b) The pressure distribution on the cornea at different time steps for IOP=15mmHg, CCT=445$\mu$m, and $\mu$=0.0683.}
\label{fig4.1}
\end{figure}

Figure \ref{fig4.1} illustrates how each individual corneal parameter affects the predicted pressure load at T = 16 ms, presenting comparisons between minimum and maximum parameter values to highlight the influence of these parameters on pressure load predictions. The predicted values aligned well with numerical data from the ABAQUS-based FSI model, as reflected in Table \ref{table3} in the Appendix, which reports RMSE values for all cases at T = 16 ms. The results indicate that peak pressure at the corneal apex varies across the two models, with higher values observed for lower IOP cases, thicker corneas, and stiffer corneas. These outcomes underscore the GBR model's capacity to accurately capture patient-specific corneal parameter variations for different cases.

\begin{table}[ht]
\centering
\begin{tabular}{| l | l | l |}
\hline
\multicolumn{2}{|l|}{Corneal parameter} & RMSE \\ \hline

\multirow{6}{*}{IOP (mmHg)} & 10 & 0.0186 \\ 
 & 13 & 0.0453 \\ 
 & 15 & 0.0511 \\ 
 & 17 & 0.0530 \\ 
 & 20 & 0.0769 \\ 
 & 25 & 0.0894 \\ \hline
 
\multirow{5}{*}{CCT (µm)} 
 & 445 & 0.0322 \\ 
 & 495 & 0.1021 \\ 
 & 545 & 0.0467 \\ 
 & 595 & 0.0458 \\ 
 & 645 & 0.0759 \\ \hline
 
\multirow{5}{*}{$\mu$} 
 & 0.0328 & 0.0201  \\ 
 & 0.0541 & 0.0340  \\ 
 & 0.0683 & 0.0119  \\ 
 & 0.0811 & 0.0362  \\ 
 & 0.1082 & 0.0533  \\ \hline

\end{tabular}
\caption{The RMSE estimated for the air puff pressure loading as obtained by the GBR algorithm by changing each corneal parameter separately.}
\label{table3}
\end{table}

\subsection{Estimation of the air pressure distribution on the cornea }	

\begin{figure}[htbp]
\centering
\includegraphics[width=0.75\textwidth]{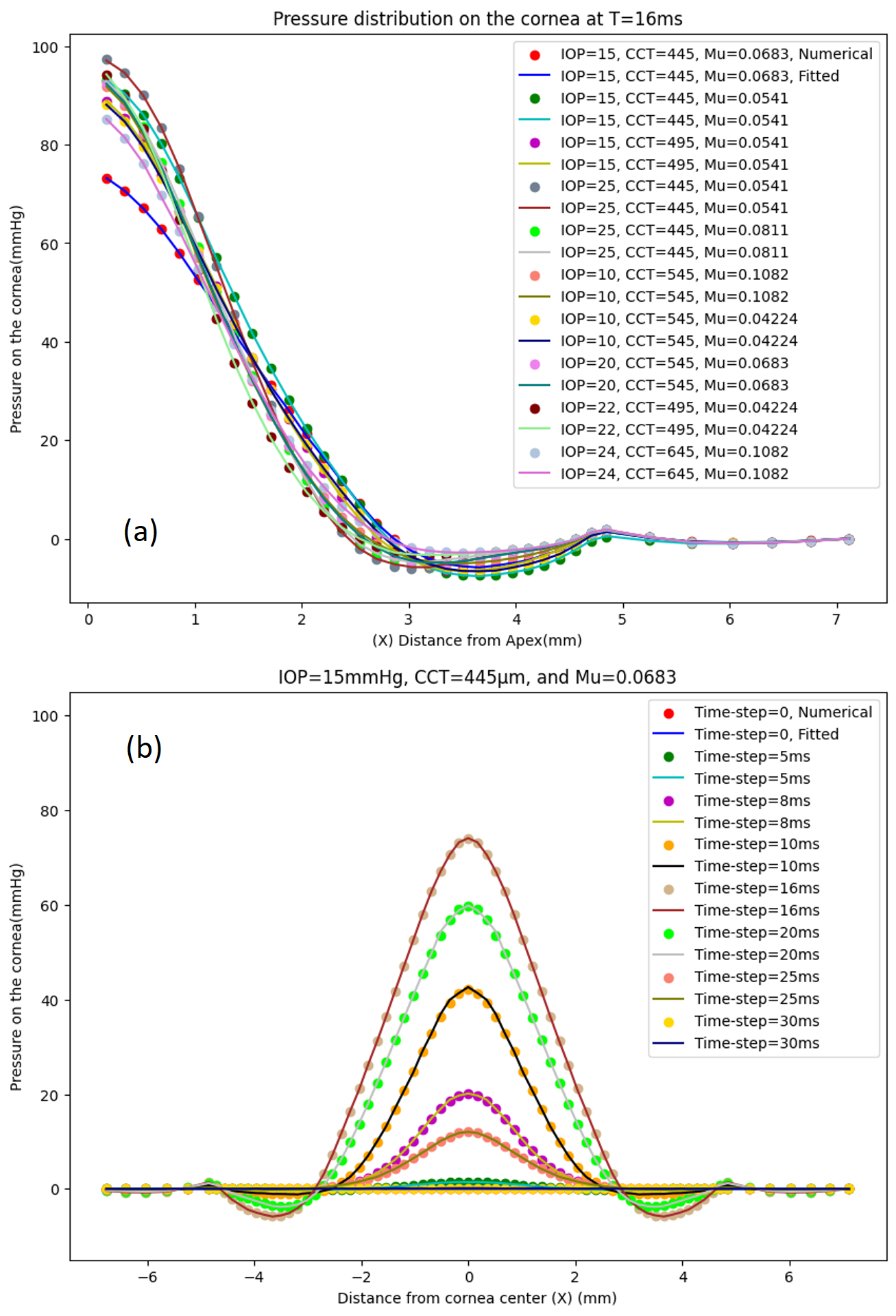}
\caption{(a)The comparison between the numerical ABAQUS model and the fitted GBR algorithm of the pressure distribution on the cornea at T=16ms with a different corneal parameter of IOP, CCT, and $\mu$, (b) The pressure distribution on the cornea at different time steps for IOP=15mmHg, CCT=445$\mu$m, and $\mu$=0.0683.}
\label{fig5.1}
\end{figure}

We analysed the impact of incorporating all corneal parameters together on the pressure loading and evaluated its performance against the numerical model. To construct our model, we simulated 17 unique cases in \texttt{ABAQUS 6-14}, using the results as input data for the algorithm. The pressure profile results, shown in Figure~\ref{fig5.1} (a), indicate that modifying a single corneal parameter can alter the entire pressure profile, while Figure~\ref{fig5.1} (b) illustrates the pressure distribution across the cornea at different time points.

The algorithm demonstrated strong alignment with the numerical model, achieving a Mean Absolute Error (MAE) of 0.0258, Root Mean Square Error (RMSE) of 0.0673, and an execution time of 93 seconds. The results reveal no consistent pattern between changes in air puff pressure loading and individual parameters, as the pressure distribution on the cornea is influenced by all parameters collectively, highlighting the importance of considering them in tandem to achieve accurate air puff pressure predictions.

\subsection{Validation of the GBR algorithm  }	

To clinically validate the proposed algorithm, a dataset containing a diverse range of corneal parameters for four healthy patients was used. Following air puff pressure predictions generated by the GBR algorithm, these predictions were applied to the finite element (FE) model of the eye in \texttt{ABAQUS 6-14}. The FE model required about ten minutes to process and produce corneal deformation results from the predicted air puff pressures. These deformation results were then compared with clinical deformation data to complete the validation process. This approach serves as an efficient alternative to the traditional CFD model by significantly reducing computational time—from approximately 28 hours (101,000 seconds) in the FSI model to just 12 minutes (720 seconds), achieving about a 99.2\% reduction in runtime.

\begin{figure}[hbt!]
\centering
\includegraphics[width=\textwidth]{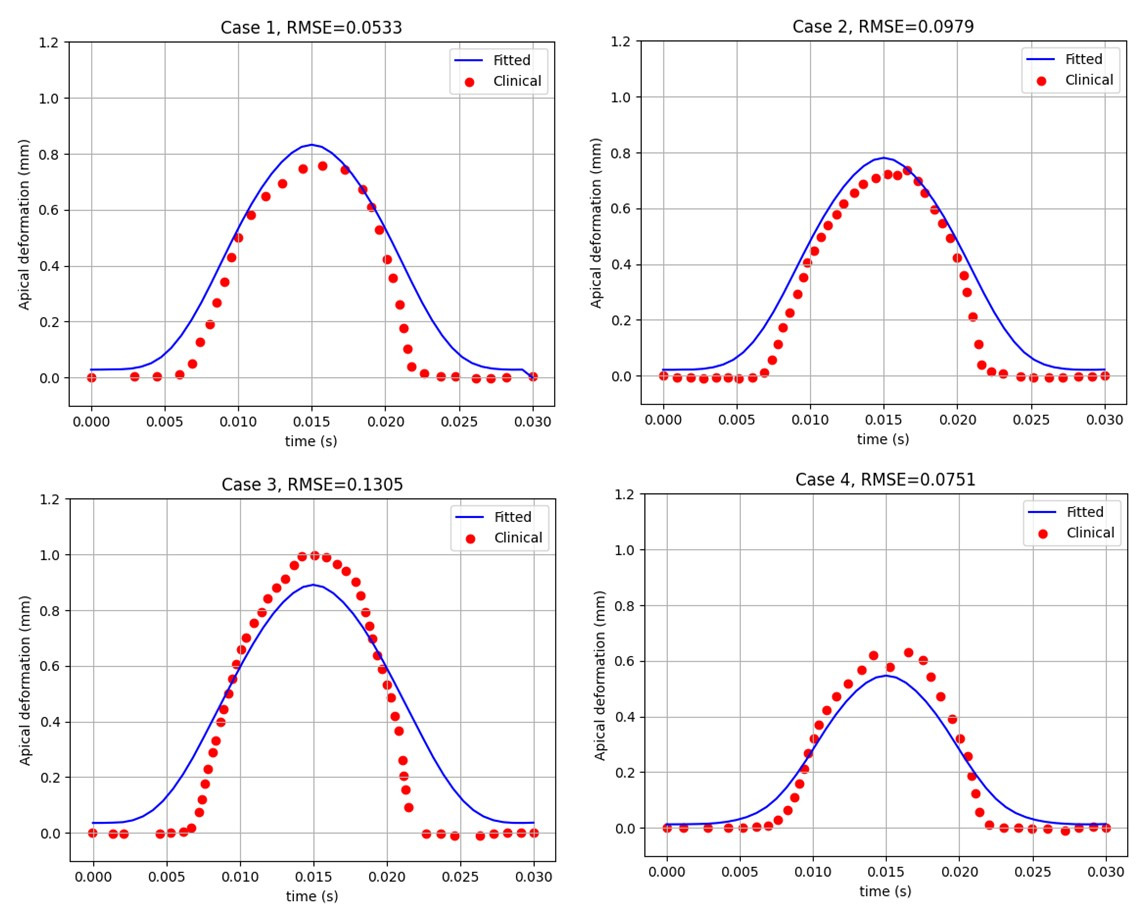}
\caption{The apical deformation resulted from the GBR+FE algorithm compared against its clinical data reference for four clinical cases.}
\label{fig7}
\end{figure}

\begin{figure}[hbt!]
\centering
\includegraphics[width=\textwidth]{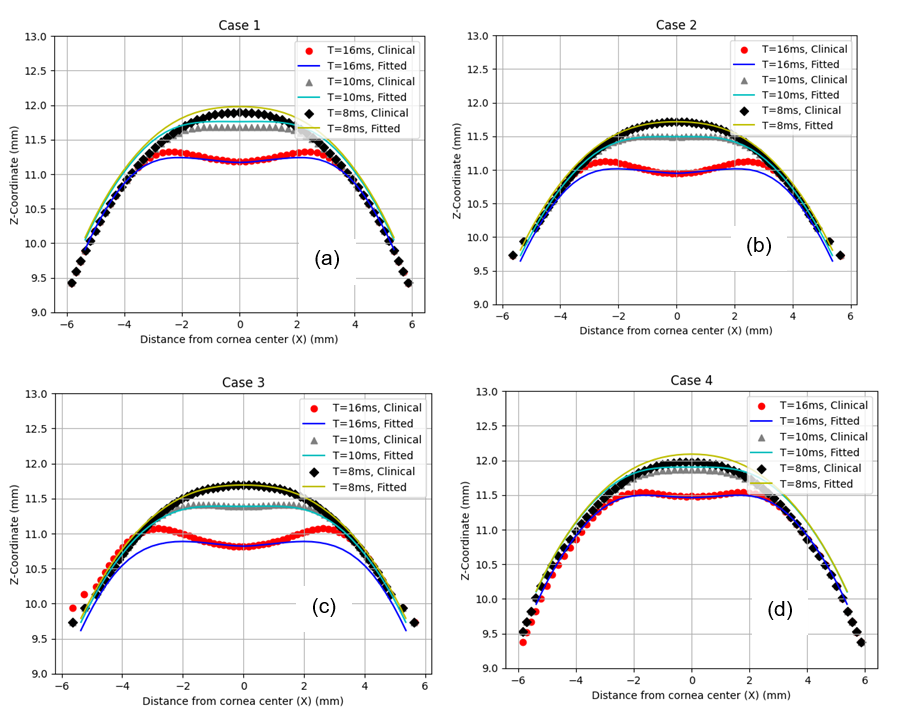}
\caption{Comparison of the corneal deformation results from the fitted algorithm with the four clinical cases}
\label{fig5}
\end{figure}

Figure ~\ref{fig7} illustrates the temporal comparison of apical deformation against clinical data, while Figure ~\ref{fig5} compares spatial corneal deformations based on the predicted air puff loading. These comparisons demonstrate that the algorithm reliably replicates actual corneal behavior. Furthermore, to assess the improvement achieved by integrating the GBR algorithm with the FE model over using the FE model alone for apical deformation predictions, RMSE values were calculated for both methods, with results presented in Table ~\ref{table4}.

\begin{table}[htp!]
\centering

\begin{tabular}{| l | l | l |}
\hline
Clinical Case & RMSE (GBR+FE model vs clinical) & RMSE (FE model only vs clinical) \\
\hline
Case 1 & 0.0533 & 0.6393 \\
\hline
Case 2 & 0.0979 & 0.3655 \\
\hline
Case 3 & 0.1305 & 0.4263 \\
\hline
Case 4 & 0.0751 & 0.4403 \\
\hline

\end{tabular}
\caption{The RMSE estimated between the clinical cases with the GBR+FE model and the numerical FE model only for the apical deformation.}
\label{table4}
\end{table}

\subsection{Results of Neural network method }

Figure \ref{fig:xt_p_dz} presents the comparison between predicted and original pressure ($P$) and deformation ($dz$) distributions on the corneal surface as functions of spatial position ($X$) and time ($t$). Each subfigure includes four 3D surfaces: the CFD original pressure ($P_{\text{original}}$, green), NN-predicted pressure ($P_{\text{predict}}$, purple), CFD original deformation ($dz_{\text{original}}$, orange), and NN-predicted deformation ($dz_{\text{predict}}$, blue). The agreement between the NN-predicted and CFD original surfaces demonstrates the accuracy of the proposed neural network in reconstructing both pressure and deformation distributions.

\begin{figure}[thp]
\begin{subfigure}{0.45\textwidth}
\includegraphics[trim={0 60 50 0},clip,width=0.9\linewidth]{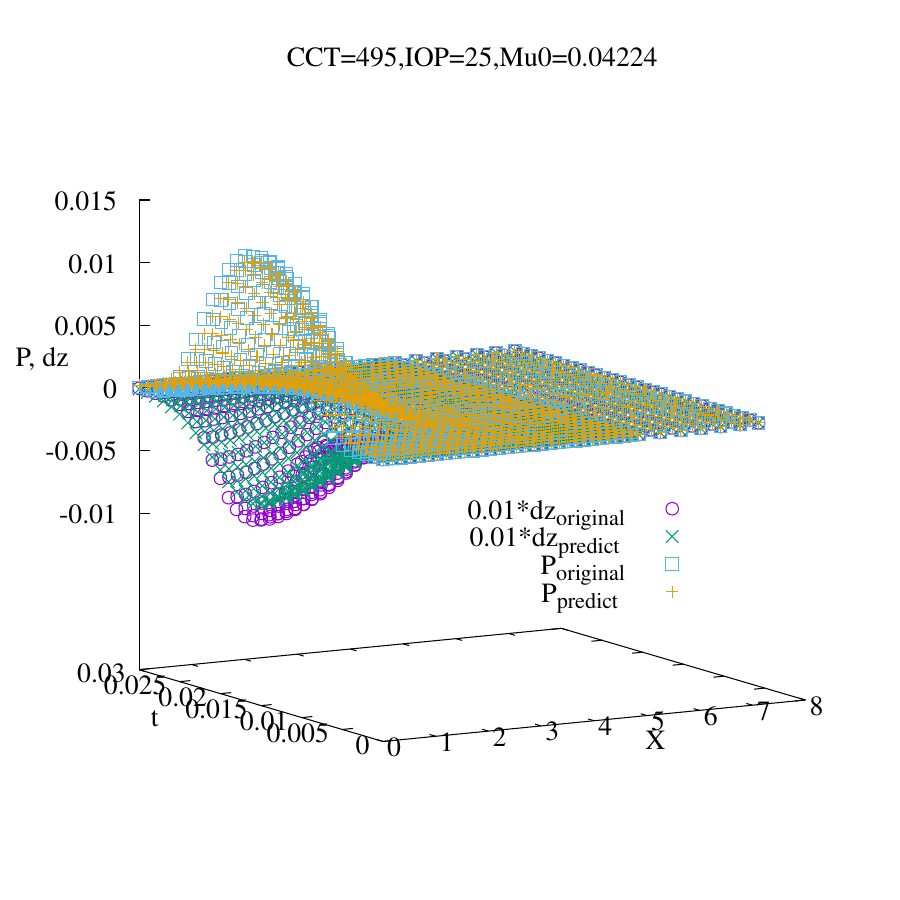}
\label{fig:case1}
\caption{ Case1}
\end{subfigure}
\begin{subfigure}{0.45\textwidth}
\includegraphics[trim={0 60 50 0},clip,width=0.9\linewidth]{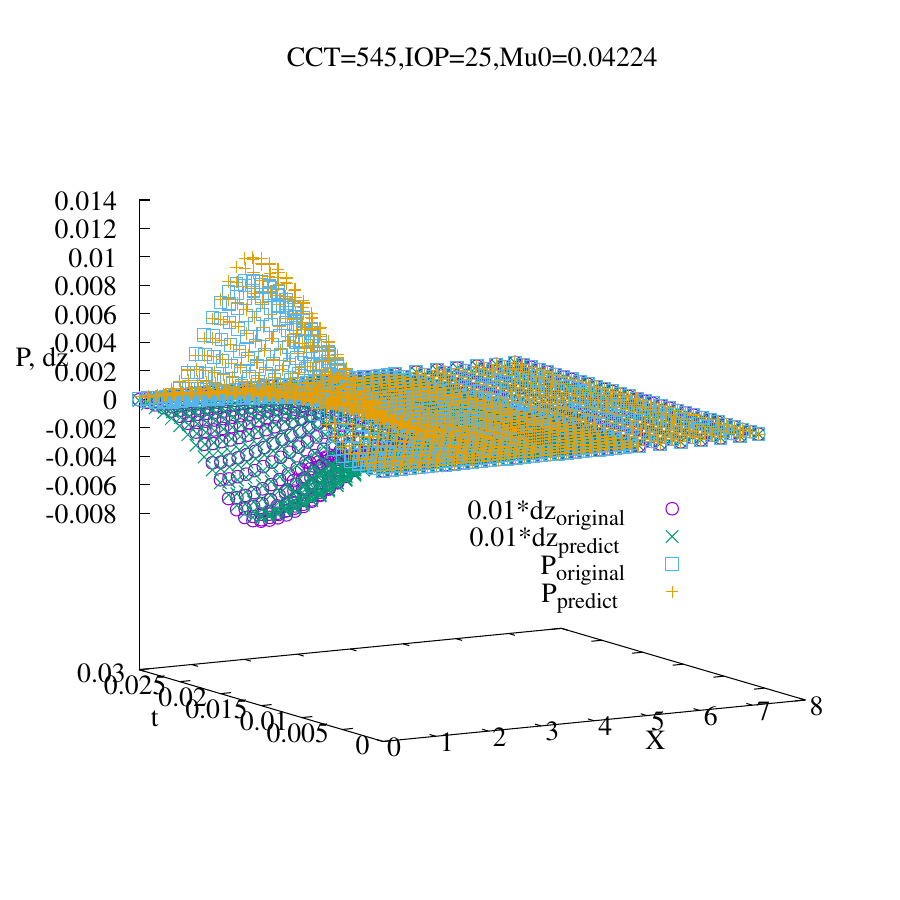}
\label{fig:case2}
\caption{Case2}
\end{subfigure}
\begin{subfigure}{0.45\textwidth}
\includegraphics[trim={0 60 50 0},clip,width=0.9\linewidth]{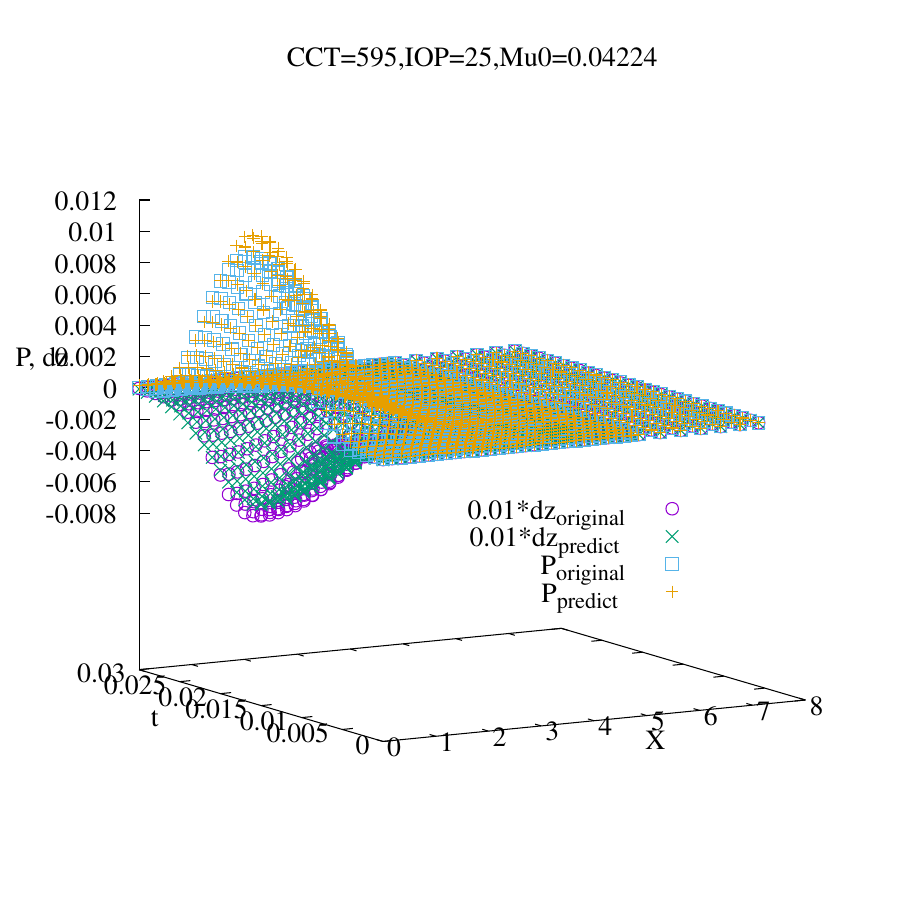}
\label{fig:case3}
\caption{Case3}
\end{subfigure}
\begin{subfigure}{0.45\textwidth}
\includegraphics[trim={0 60 50 0},clip,width=0.9\linewidth]{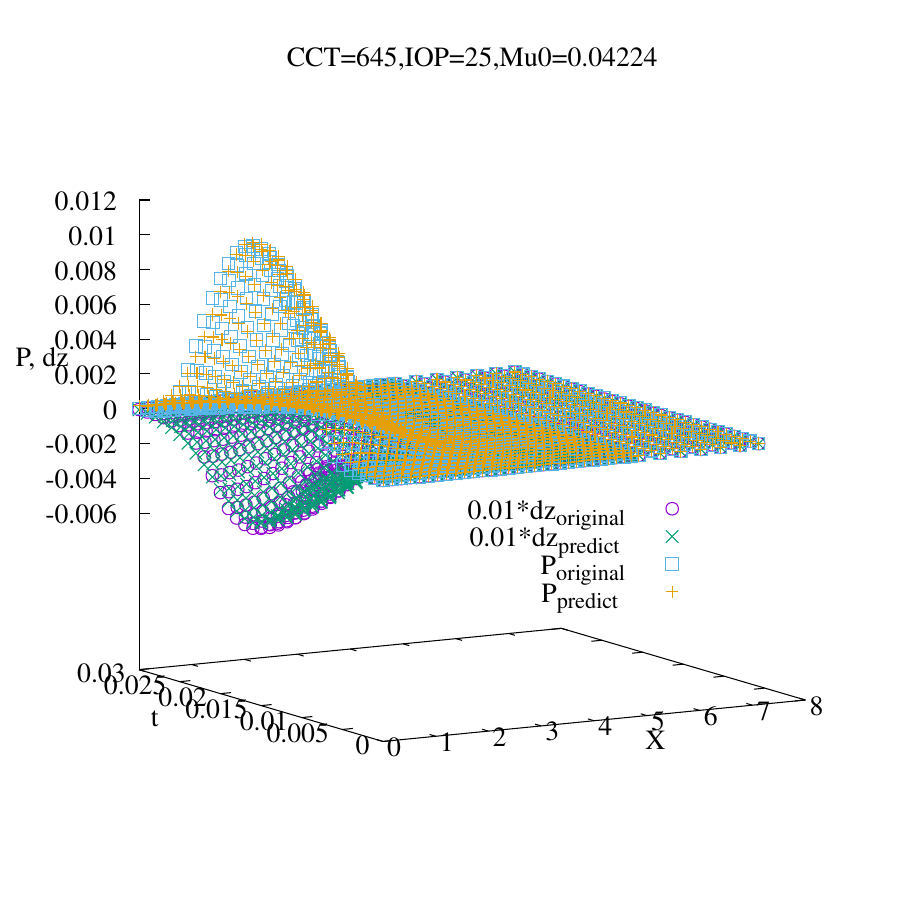}
\label{fig:case4}
\caption{Case4}
\end{subfigure}
\begin{subfigure}{0.45\textwidth}
\includegraphics[trim={0 60 50 0},clip,width=0.9\linewidth]{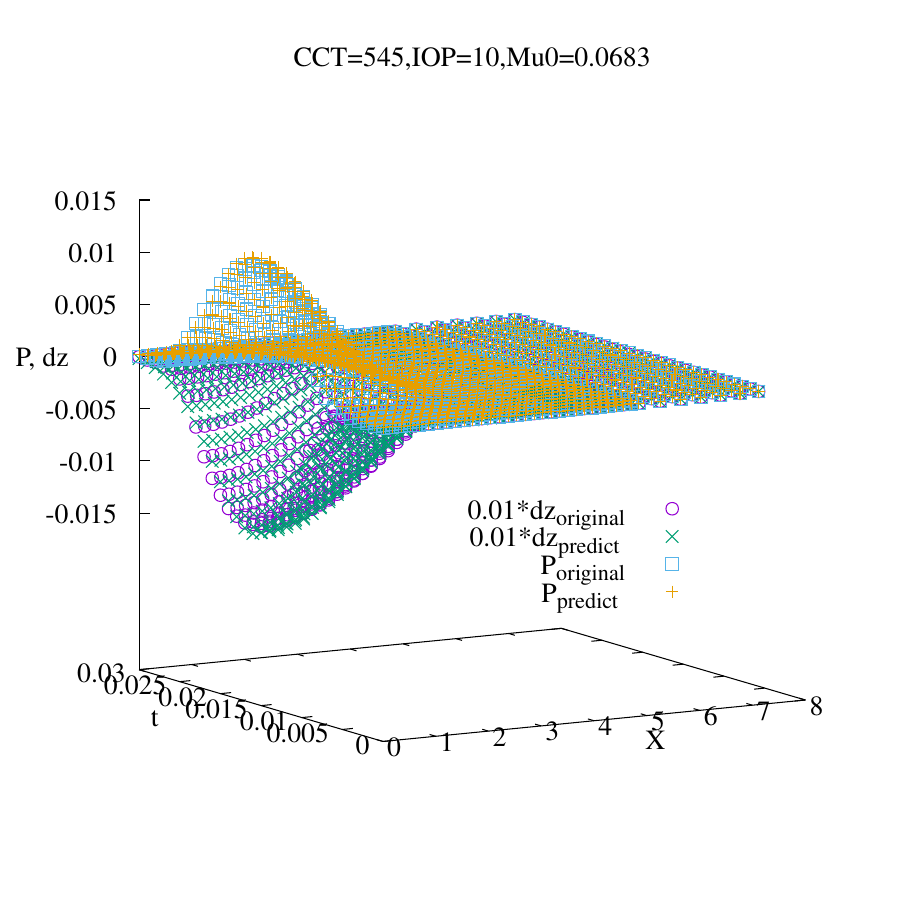}
\label{fig:case5}
\caption{Case5}
\end{subfigure}
\begin{subfigure}{0.45\textwidth}
\includegraphics[trim={0 60 50 0},clip,width=0.9\linewidth]{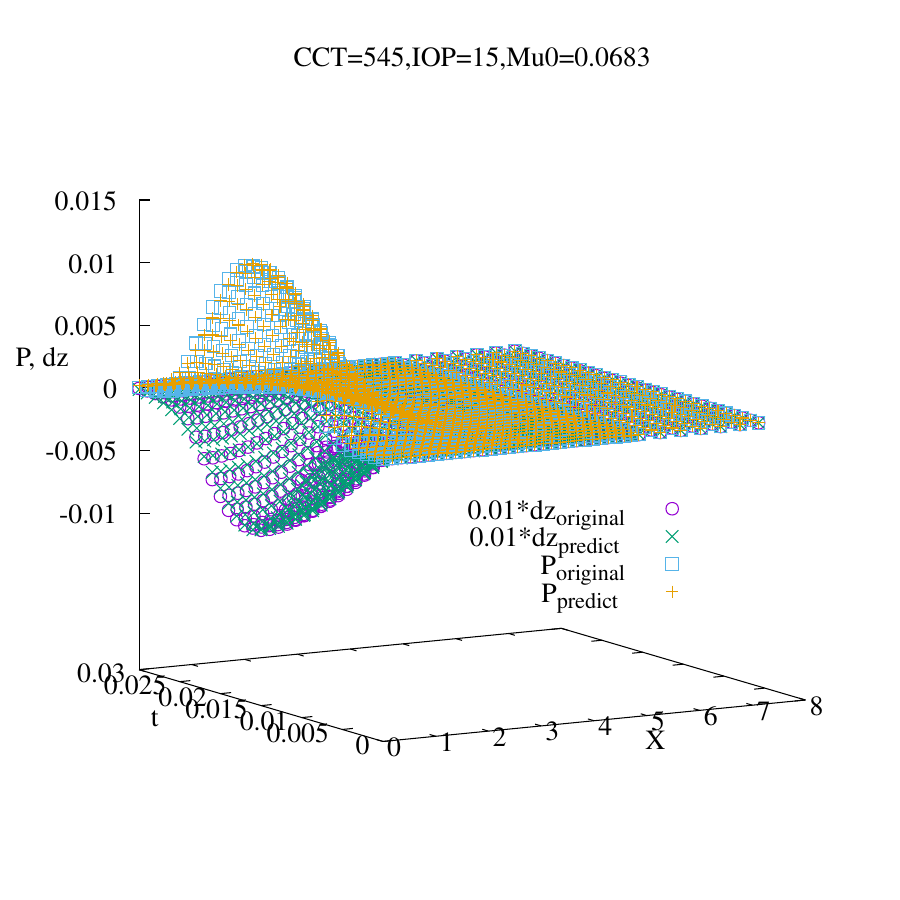}
\label{fig:case6}
\caption{Case6}
\end{subfigure}
\end{figure}

\begin{figure}[p] 
    \ContinuedFloat
\begin{subfigure}{0.45\textwidth}
\includegraphics[trim={0 60 50 0},clip,width=0.9\linewidth]{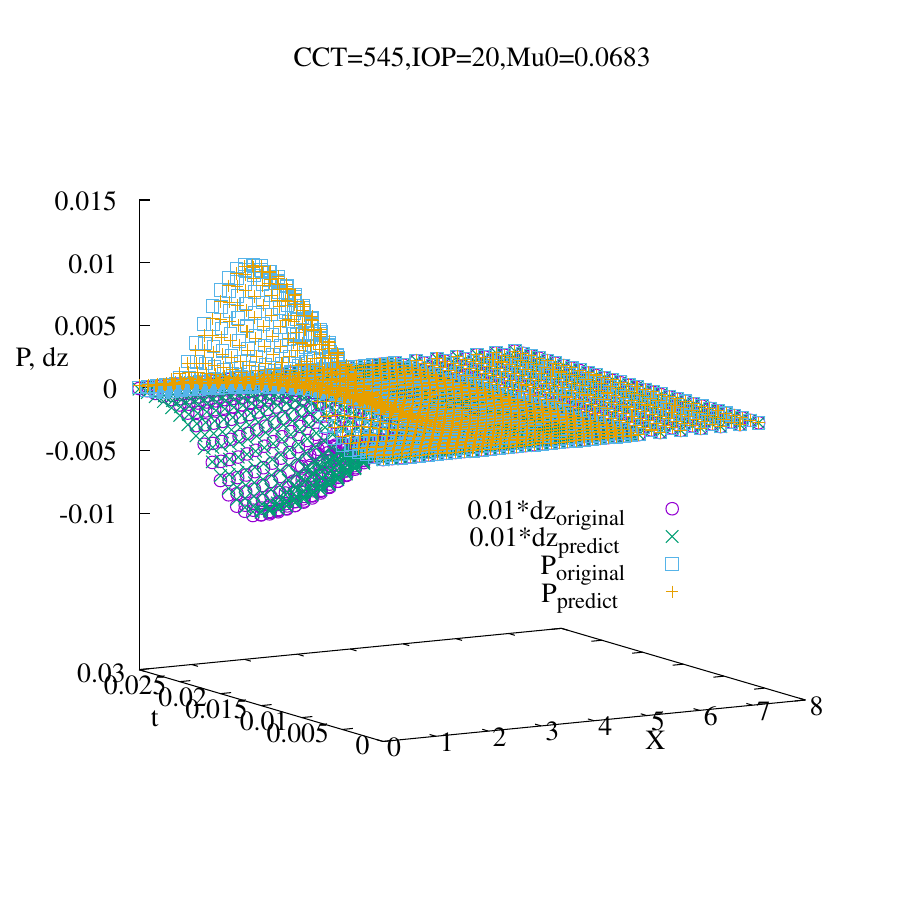}
\label{fig:case7}
\caption{Case7}
\end{subfigure}
\begin{subfigure}{0.45\textwidth}
\includegraphics[trim={0 60 50 0},clip,width=0.9\linewidth]{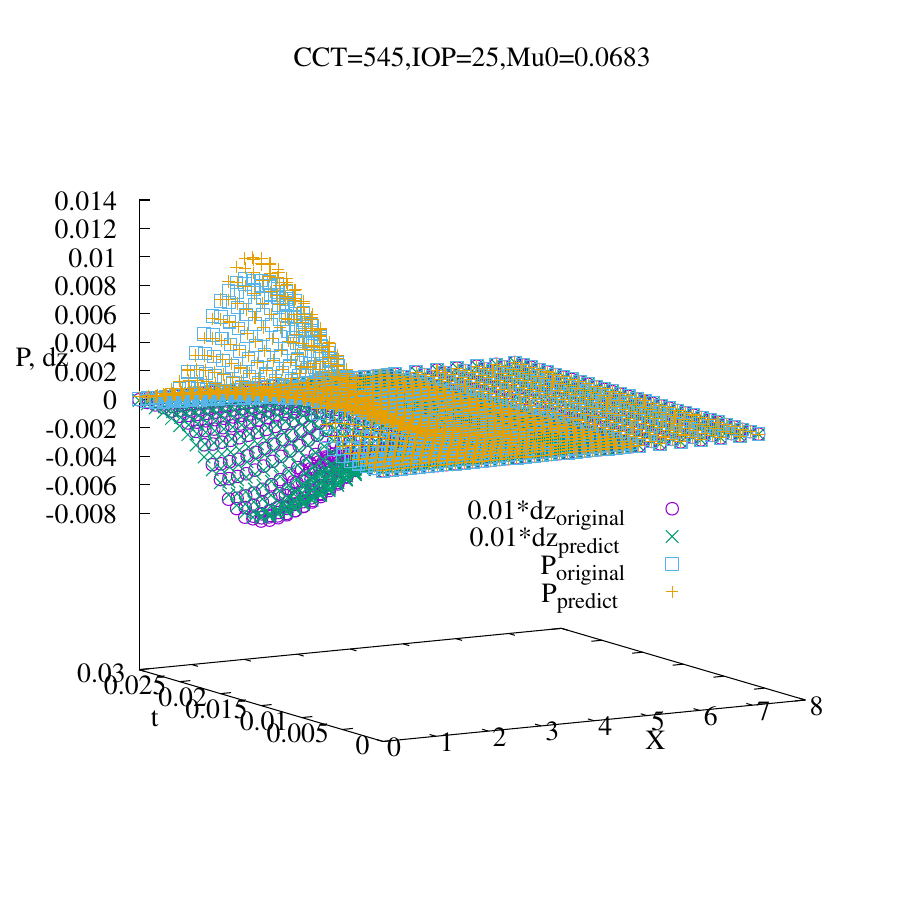}
\label{fig:case8}
\caption{Case8}
\end{subfigure}
\begin{subfigure}{0.45\textwidth}
\includegraphics[trim={0 60 50 0},clip,width=0.9\linewidth]{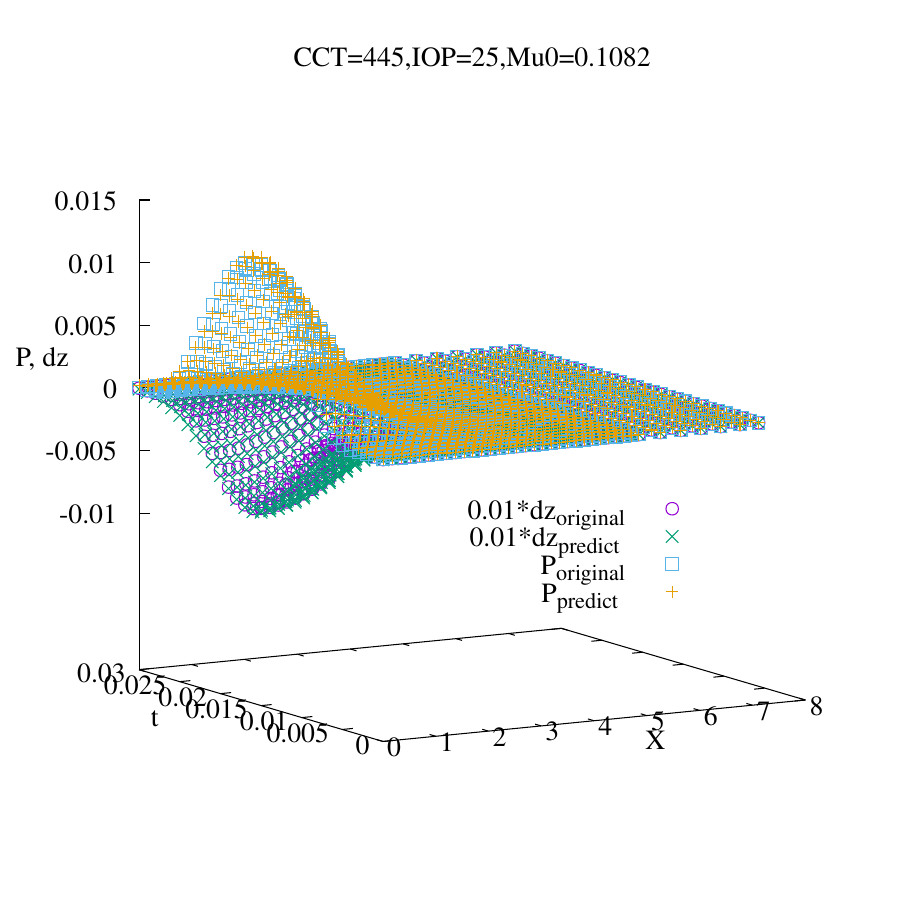}
\label{fig:case9}
\caption{Case9}
\end{subfigure}
\begin{subfigure}{0.45\textwidth}
\includegraphics[trim={0 60 50 0},clip,width=0.9\linewidth]{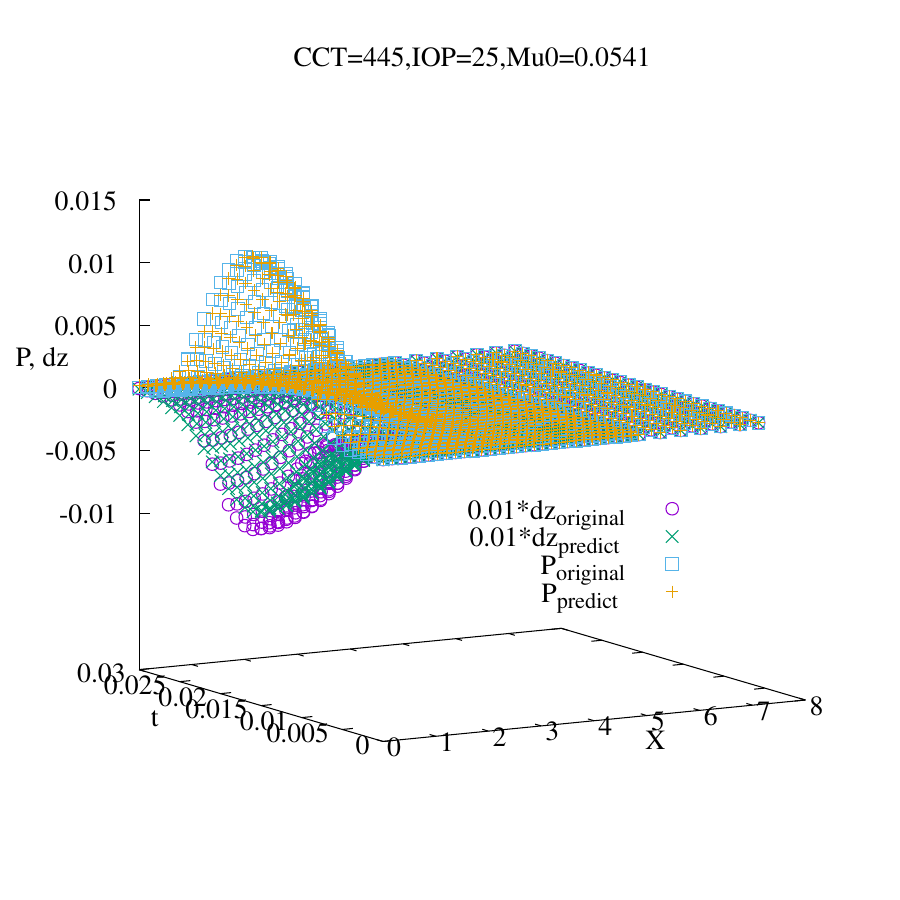}
\label{fig:case11}
\caption{Case11}
\end{subfigure}
\begin{subfigure}{0.45\textwidth}
\includegraphics[trim={0 60 50 0},clip,width=0.9\linewidth]{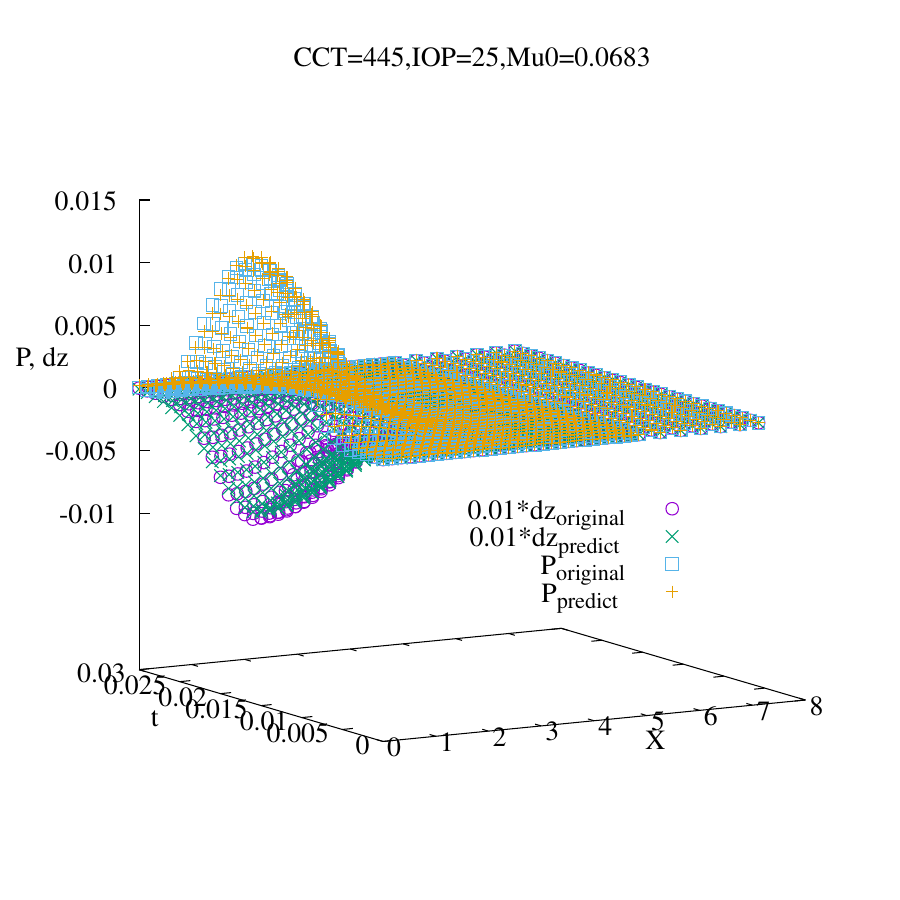}
\label{fig:case12}
\caption{Case12}
\end{subfigure}
\begin{subfigure}{0.45\textwidth}
\includegraphics[trim={0 60 50 0},clip,width=0.9\linewidth]{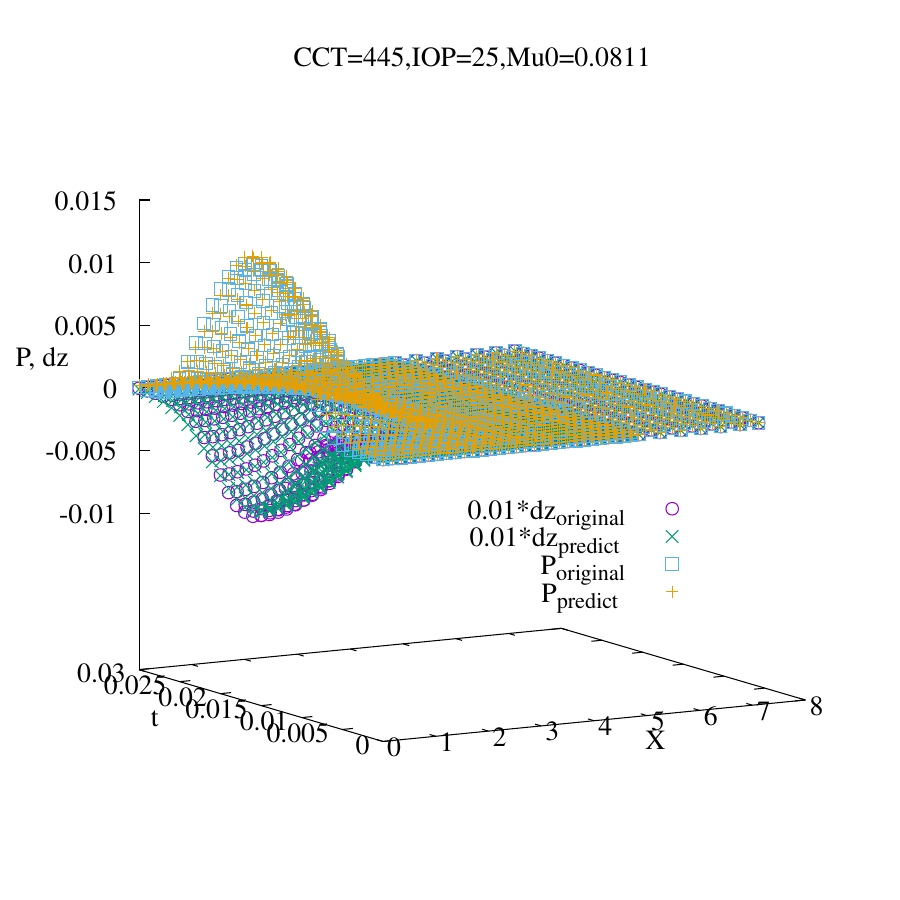}
\label{fig:case13}
\caption{Case13}
\end{subfigure}
\caption{The predicted pressure $P$ and deformation $dz$ on the cornea location $X$ at time $t$. Green: CFD pressure $P_{original}$, Purple: NN predicted pressure $P_{predict}$, Orange: CFD deformation $dz_{original}$, Blue: NN predicted deformation $dz_{predict}$ }
\label{fig:xt_p_dz}
\end{figure}

\begin{figure}[h!]
\begin{subfigure}{0.3\textwidth}
\includegraphics[trim={0 0 10 0},clip,width=1\linewidth]{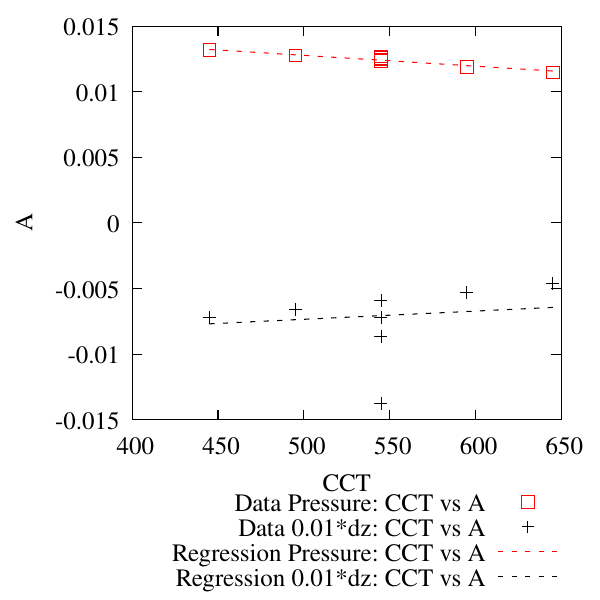}
\label{fig:CCT-A}
\caption{ $CCT-A$}
\end{subfigure}
\begin{subfigure}{0.3\textwidth}
\includegraphics[trim={0 0 10 0},clip,width=1\linewidth]{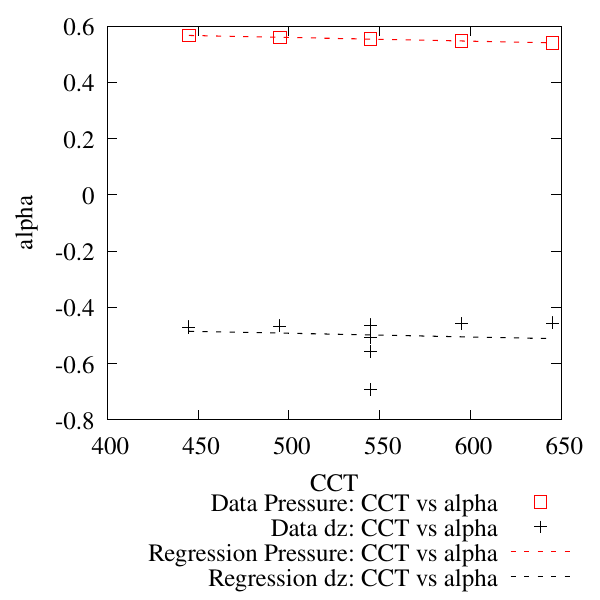}
\label{fig:CCT-alpha}
\caption{$CCT-\alpha$}
\end{subfigure}
\begin{subfigure}{0.3\textwidth}
\includegraphics[trim={0 0 10 0},clip,width=1\linewidth]{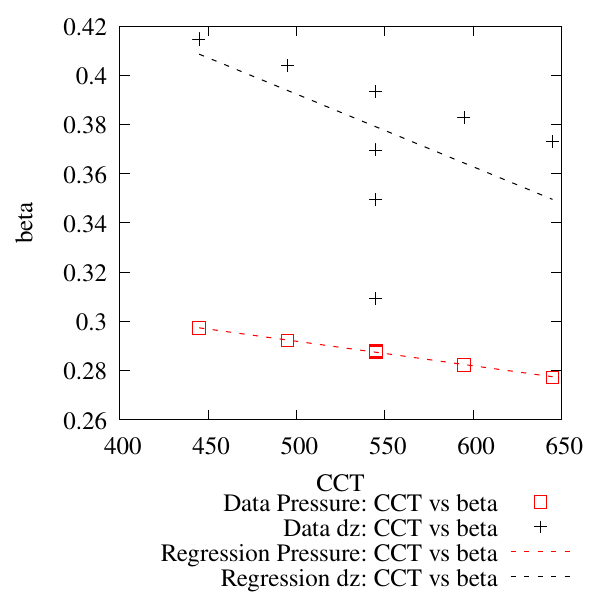}
\label{fig:CCT-beta}
\caption{$CCT-\beta$}
\end{subfigure}

\begin{subfigure}{0.3\textwidth}
\includegraphics[trim={0 0 10 0},clip,width=1\linewidth]{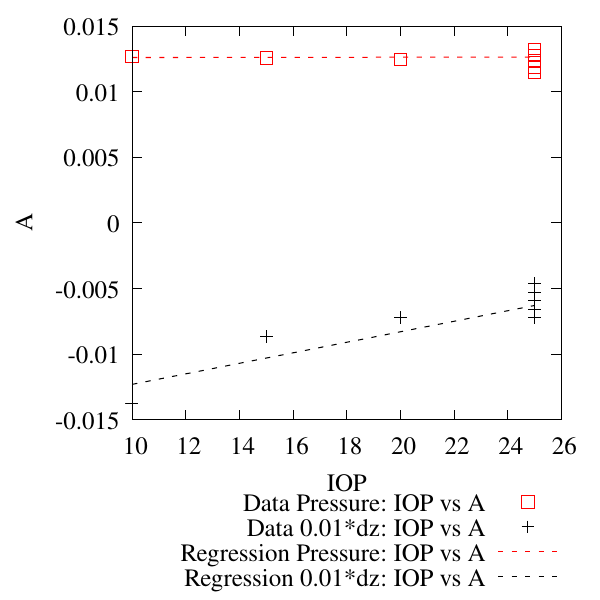}
\label{fig:IOP-A}
\caption{ $IOP-A$}
\end{subfigure}
\begin{subfigure}{0.3\textwidth}
\includegraphics[trim={0 0 10 0},clip,width=1\linewidth]{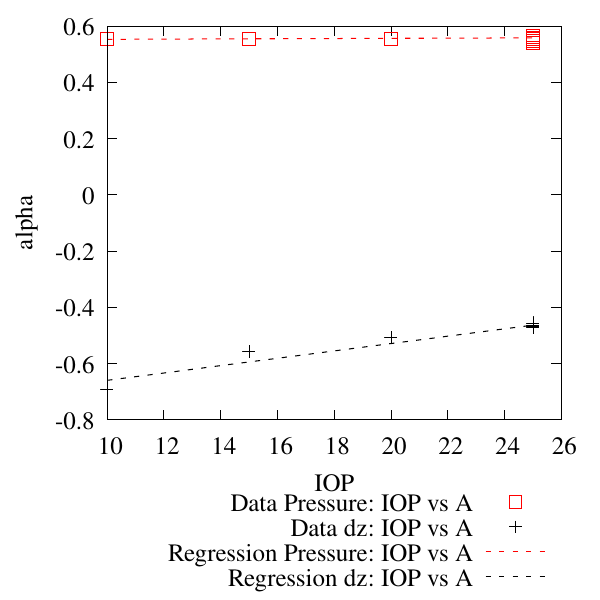}
\label{fig:IOP-alpha}
\caption{$IOP-\alpha$}
\end{subfigure}
\begin{subfigure}{0.3\textwidth}
\includegraphics[trim={0 0 10 0},clip,width=1\linewidth]{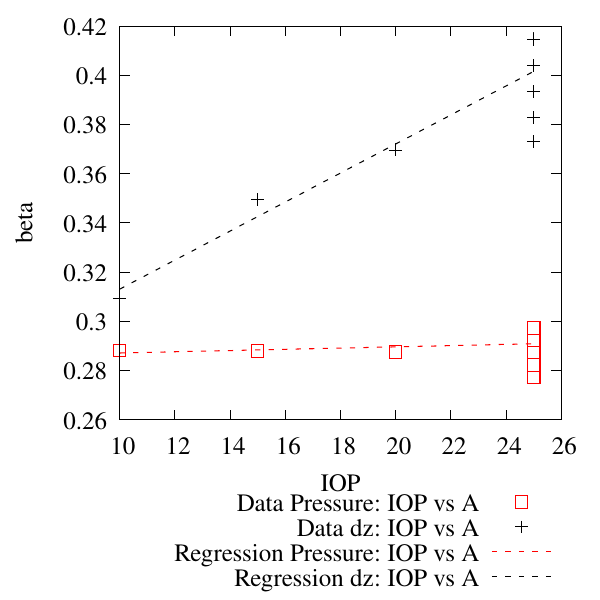}
\label{fig:IOP-beta}
\caption{$IOP-\beta$}
\end{subfigure}

\begin{subfigure}{0.3\textwidth}
\includegraphics[trim={0 0 10 0},clip,width=1\linewidth]{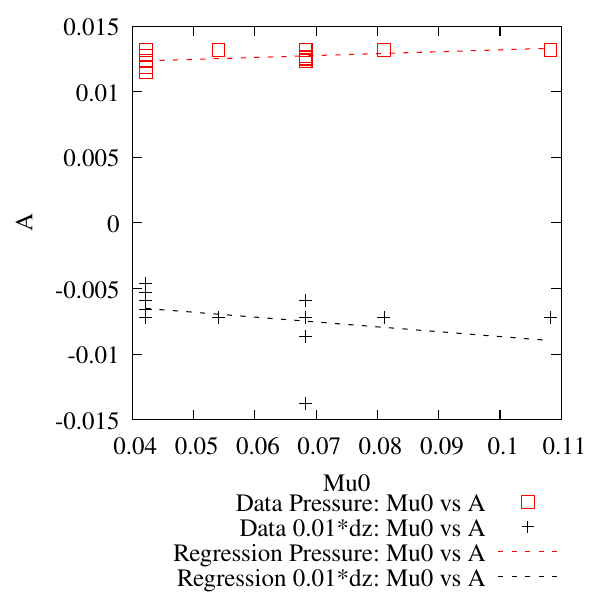}
\label{fig:Mu0-A}
\caption{ $Mu0-A$}
\end{subfigure}
\begin{subfigure}{0.3\textwidth}
\includegraphics[trim={0 0 10 0},clip,width=1\linewidth]{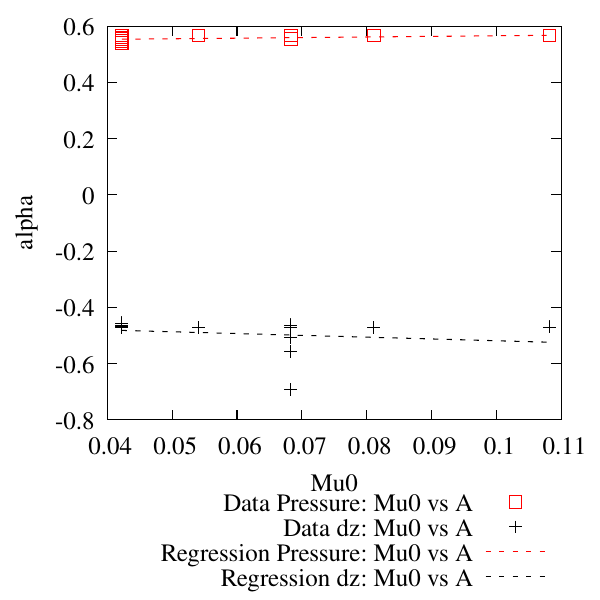}
\label{fig:Mu0-alpha}
\caption{$Mu0-\alpha$}
\end{subfigure}
\begin{subfigure}{0.3\textwidth}
\includegraphics[trim={0 0 10 0},clip,width=1\linewidth]{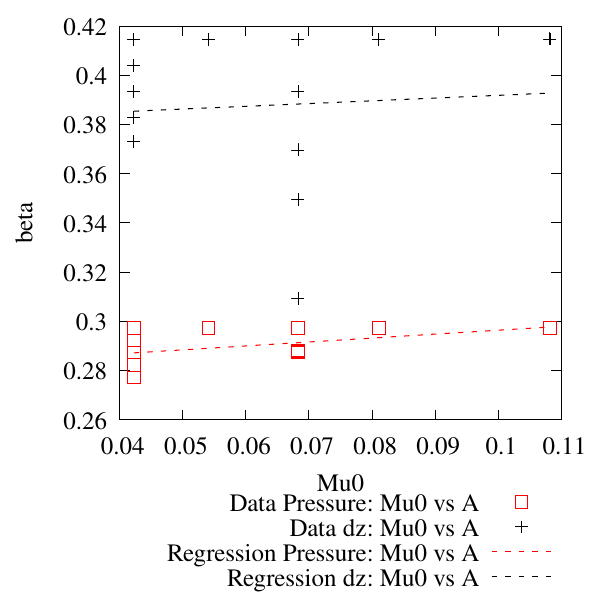}
\label{fig:Mu0-beta}
\caption{$Mu0-\beta$}
\end{subfigure}

\caption{Correlation between input parameters $CCT$, $IOP$, $Mu0$ and output coefficients in the reconstructed relation $A$, $\alpha$, $\beta$ }
\label{fig:corr}
\end{figure}

The results reveal that the temporal variations of pressure and deformation are similar, as indicated by their coefficients $\mu$ (temporal mean) and $\sigma$ (temporal standard deviation). This similarity suggests shared temporal dynamics for pressure and cornea deformation. However, significant differences are observed in the spatial patterns, governed by coefficients $A$, $\alpha$, and $\beta$. Specifically, $A$ for deformation has an opposite sign and a smaller magnitude compared to $A$ for pressure, with $A_{\text{deformation}} \approx -0.01 \cdot A_{\text{pressure}}$. These distinctions highlight the different spatial attenuation and scaling behaviors of pressure and deformation, with deformation exhibiting less steeper spatial variability. Figure~1 demonstrates that the NN with the proposed wavelet function accurately and efficiently captures both shared temporal and distinct spatial characteristics of pressure and deformation.

Figure \ref{fig:corr} illustrates the correlations between the input parameters ($\text{CCT}$, $\text{IOP}$, $\mu_0$) and the output coefficients ($A$, $\alpha$, $\beta$, $\mu$, $\sigma$) for both pressure and deformation. Each subplot shows scatterplots with fitting lines to describe the correlation.

The analysis indicates that pressure and deformation exhibit different dependencies on the input parameters. For instance, $A_{\text{pressure}}$ correlates positively with IOP, while $A_{\text{deformation}}$ shows a negative correlation with CCT. Temporal coefficients $\mu$ and $\sigma$ demonstrate minor correlations with all inputs, reflecting shared temporal influences between pressure and deformation. In contrast, spatial coefficients $\alpha$ and $\beta$ exhibit high sensitivity to  inputs $CCT$, $IOP$ and slightly lower to $\mu_0$, with $\beta_{\text{pressure}}$ strongly influenced by $CCT$, whereas $\beta_{\text{deformation}}$ depends more on $CCT$ and $IOP$. This parametric modeling shows consistency for pressure and deformation.
Figure \ref{fig:corr} demonstrates the relationships between corneal fluid-solid interaction and the observed pressure and deformation distributions.

\section{Conclusion}
Estimating the air puff pressure distribution on the cornea is crucial for enhancing the accuracy of intraocular pressure (IOP) measurements, which are vital for detecting and assessing corneal diseases. The precision of these measurements directly impacts the assessment of corneal material properties. While the Fluid-Structure Interaction (FSI) method has shown high accuracy in predicting corneal compressive loads based on deformation, it requires extensive computational time, often up to 28 hours. This study introduces a supervised machine learning (ML) regression algorithm designed to predict corneal pressure distribution from specific corneal parameters while optimizing accuracy and reducing computational demand. 

The primary outcome is a practical Gradient Boosting Regressor (GBR) model that effectively estimates corneal pressure distribution by accounting for key parameters: IOP, central corneal thickness (CCT), the corneal material coefficient (indicative of age), and the time step during testing. Our results reveal that air puff pressure loading is significantly influenced by the intricate variations in corneal parameters unique to each patient. Additionally, the algorithm substantially decreases computational time from approximately 101,000 seconds (28 hours) to 720 seconds (12 minutes)—a reduction of about 99.2\%—while maintaining the accuracy offered by the FSI model.

This advancement in computational efficiency could greatly enhance the development of parametric equations for IOP and the Stress-Strain Index (SSI) by enabling analyses of a larger number of fully individualized eye models with dynamic topography, which will be the focus of our future research. Additionally, we propose incorporating Physics-Informed Neural Networks (PINNs) into the CFD-based air puff model as part of our ongoing efforts. PINNs can leverage governing physical equations within the ML framework to provide faster and more accurate simulations, further reducing computational costs while enhancing the predictive capabilities of air puff-based tonometry models. This innovative integration of PINNs has the potential to accelerate clinical workflows and improve the reliability of tonometry measurements in patient-specific applications. 

This study introduces a neural network framework to model the pressure and deformation distributions on the corneal surface during air puff. By learning the relationships between input parameters (CCT, IOP, and $\mu_0$) and five governing coefficients, the NN efficiently reconstructs high-fidelity CFD data. The results demonstrate the network’s ability to capture shared temporal dynamics and distinct spatial patterns between pressure and deformation. Correlation analyses reveal dependencies of pressure and deformation on biomechanical inputs, showing the tailored parametric models. These findings establish the NN framework as a computationally efficient and interpretable tool for analyzing fluid-structure interaction in ophthalmology. Future work will extend this framework to incorporate patient-specific geometries and dynamic boundary conditions for enhanced predictive capability.

\section*{Acknowledgments}
The authors would like to thank the Leeds Institute for Fluid Dynamics for hosting the National Fellowships in Fluid Dynamics (NFFDY) 2024 summer programme funded by EPSRC. Special thanks to Prof. Steve Tobias and Dr. Claire Savy for the programme arrangement and organisation.

\bibliography{sample}

\end{document}